\documentclass[graybox,onecollarge]{svmult}

\usepackage{mathptmx}       
\usepackage{helvet}         
\usepackage{courier}        
                            
\usepackage{makeidx}         
\usepackage{graphicx}        
\usepackage[bottom]{footmisc}

\usepackage{url}

\begin{document}

\title*{The environment of radio sources in the VLA-COSMOS Survey field}

\author{Nicola Malavasi \and Sandro Bardelli \and Paolo Ciliegi \and Olivier Ilbert \and Lucia Pozzetti \and Elena Zucca}
\authorrunning{Nicola Malavasi et al.} 

\institute{Nicola Malavasi \at University of Bologna, Department of Physics and Astronomy (DIFA), v.le Berti Pichat 6/2 - 40127 Bologna, Italy, \email{nicola.malavasi@unibo.it}
\and Sandro Bardelli \and Paolo Ciliegi \and Lucia Pozzetti \and Elena Zucca \at INAF--Osservatorio Astronomico di Bologna, via Ranzani 1 - 40127 Bologna, Italy \and Olivier Ilbert \at Aix Marseille Universit\'{e}, CNRS, LAM (Laboratoire d'Astrophysique de Marseille), UMR 7326, 13388 Marseille, France}

\maketitle

\abstract{This work studies the correlation among environmental density and radio AGN presence up to $z = 2$. Using data from the photometric COSMOS survey and its radio 1.4 GHz follow-up (VLA-COSMOS), a sample of radio AGNs has been defined. The environment was studied using the richness distributions inside a parallelepiped with base side of 1 Mpc and height proportional to the photometric redshift precision. Radio AGNs are found to be always located in environments significantly richer than those around galaxies with no radio emission. Moreover, a distinction based on radio AGN power shows that the significance of the environmental effect is only maintained for low-power radio sources. The results of this work show that denser environments play a significant role in enhancing the probability that a galaxy hosts a radio AGN and, in particular, low-power ones.}

\section{Introduction}
\label{intro}
The problem of the transformation of the galaxy population from star-forming to quiescent is still an open one in modern astrophysics. General agreement has been reached on the fact that galaxy mass, galaxy environment and AGN feedback play a major role in star formation quenching. It has been suggested (see \cite{hickox09}) that the central AGN co-evolves with the host-galaxy: while the host-galaxy transforms from a star-forming to a quiescent one, the AGN passes from a quasar, X-ray emitter phase to a radio-galaxy one. These transformations happen at earlier epochs for haloes of higher mass, that were found to reside primarily in high-density environments, where early-type galaxies dominate at low redshifts (\cite{quadri12,chuter11}). Moreover, it was already known that many radio AGNs reside in early-type galaxies (\cite{ledlow96}), that the probability that a galaxy hosts a radio AGN is increasing with stellar mass (\cite{bardelli09}), and that the fraction of radio active early-type galaxies is an increasing function of local density (\cite{bardelli10}). In this work (see \cite{malavasi15}) the environment of radio sources of the VLA-COSMOS survey (\cite{schinnerer07}), cross-identified with the COSMOS photometric redshift sample (\cite{ilbert09}), is explored.

\section{Data And Method}
\label{data}
The analysis was performed on the environment of a selection of 272 radio AGNs from the VLA-COSMOS survey (\cite{schinnerer07,schinnerer10}). This survey (which is composed of 1.4 GHz data, with a sensitivity of about 11 $\mu$Jy r.m.s.) was cross-correlated with the COSMOS photometric survey (\cite{scoville07}) with photometric redshifts measured by \cite{ilbert09}. The COSMOS survey catalogue is composed of optical galaxies down to $i^+ < 26.5$, which were used both as tracers for the environment around AGN sources and as extraction pool for the control samples. The accuracy of the photometric redshifts ($z_p$) is estimated to be $\sigma_{\Delta z/(1+z)}  = 0.06$.

\begin{figure}
\begin{minipage}{0.5\textwidth}
\resizebox{\hsize}{!}{\includegraphics{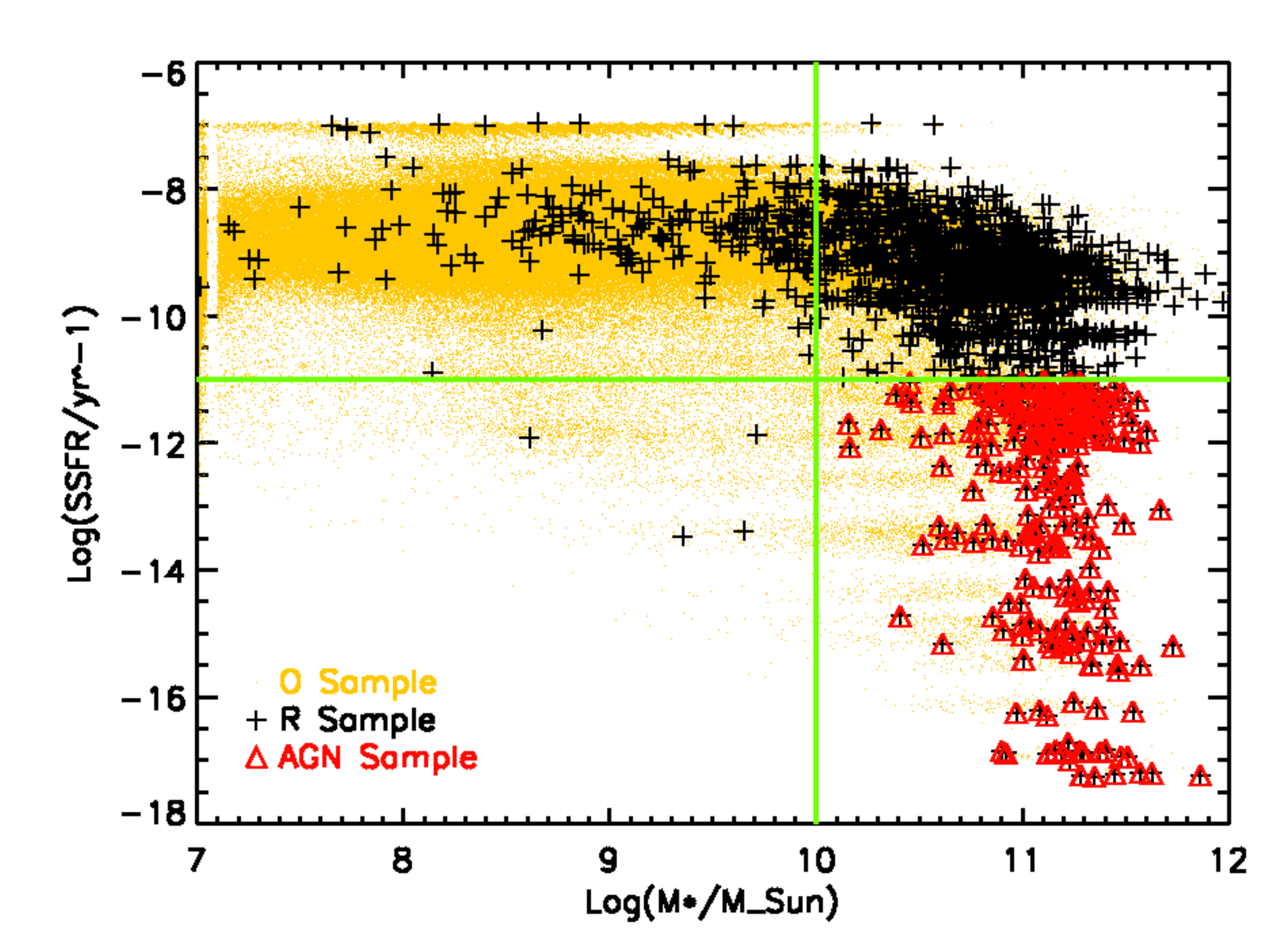}}
\end{minipage}
\begin{minipage}{0.5\textwidth}
\resizebox{\hsize}{!}{\includegraphics{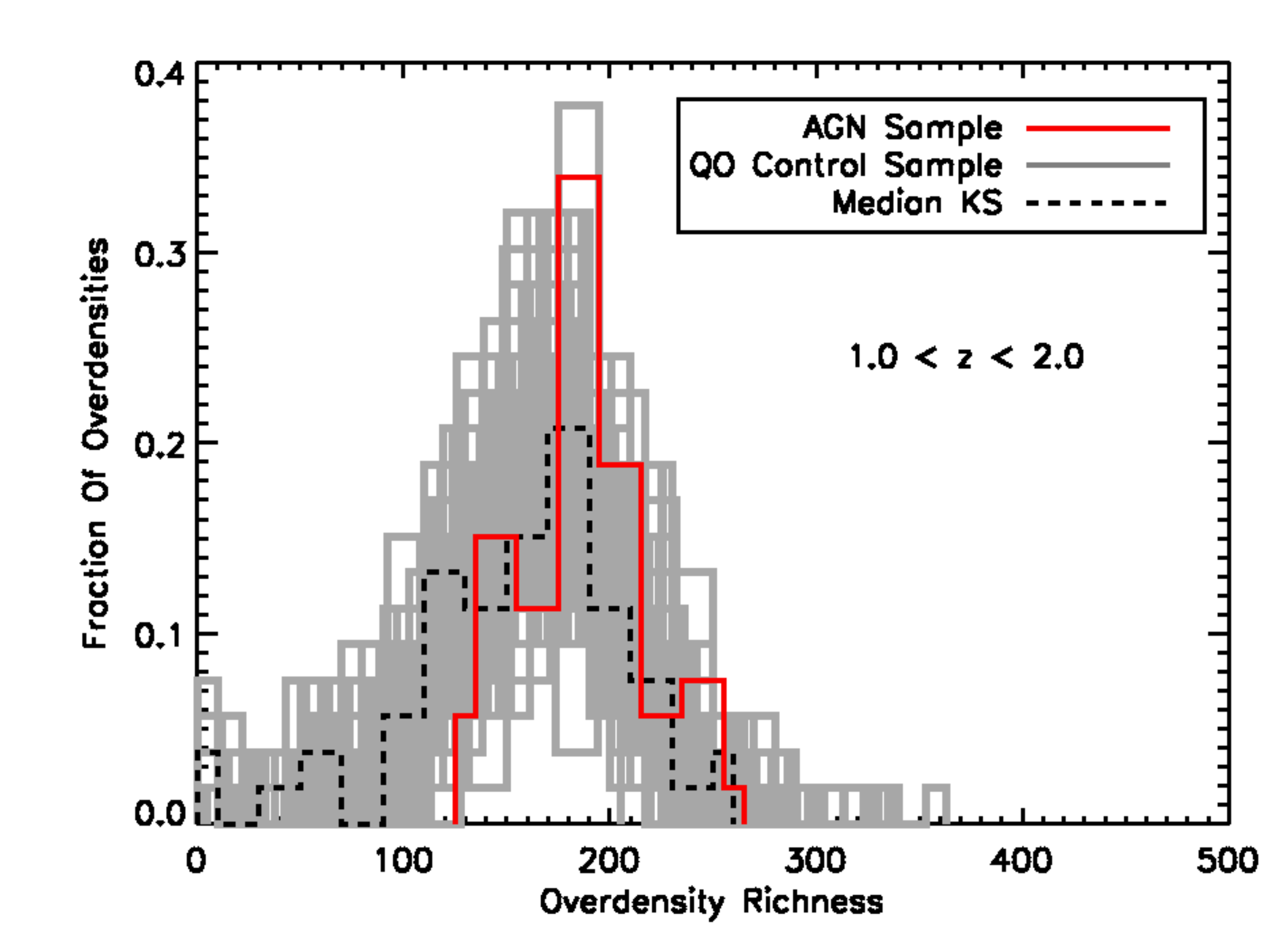}}
\end{minipage}
\caption{Left: \textit{SSFR - $M^{\ast}$ plane}. The black crosses refer to the full sample of VLA-COSMOS radio sources, the red triangles to the radio AGN sample, and the green lines to the cuts in stellar mass and SSFR described in text. In yellow, the full COSMOS optical sample is reported for comparison. Right: \textit{Galaxy overdensity richness distribution, samples AGN and QO}. This plot refers to $1 \le z \le 2$. The solid red line represents the AGN sample, the grey lines the 100 extractions of the QO control sample. The dashed black line is the richness distribution of the control sample extraction corresponding to the median value of the KS probability value distribution.}
\label{figure}
\end{figure}

AGNs were extracted among radio sources as those hosted by massive and quiescent galaxies through a cut to $\log(M^{\ast}/M_{\odot}) \ge 10$ and  $\log(SSFR/yr^{-1}) \le -11$, as shown in the left panel of Figure \ref{figure}. The control sample of normal galaxies QO has been extracted from the same lower-right region of the SSFR - $M^{\ast}$ plane. In order to have a fully representative control sample, galaxies were randomly extracted with the same mass distribution of the radio AGNs.

The environment has been estimated around every AGN source and control galaxy by counting optical galaxies in a parallelepiped with a base side of 1 Mpc (comoving) and height $2 \cdot \varDelta z = 2 \times 3 \times \sigma_{\varDelta z/(1+z)} \times (1+z_p)$, in three different redshift bins: $z \in [0.0-0.7[$, $[0.7-1.0[$ and $[1.0-2.0]$. The number of radio AGNs is respectively 119 sources, 100 sources, and 53 sources.

\section{Results And Conclusions}
\label{results}
It was found that the environment around radio AGNs is significantly denser than the environment around sources from the control sample (\emph{i.e.} that show no sign of radio emission). This is visible in the right panel of Figure \ref{figure}, which shows the overdensity richness distribution for the AGN sample and for the 100 independent extractions of the control sample in the farthest redshift bin. A Kolmogorov-Smirnov test between the distributions results in median values of the KS test probability value distribution of $8.6 \times 10^{-5}$, $6.0 \times 10^{-6}$, and 0.006 in each redshift bin respectively.

The AGN sample has been further divided according to its radio power: a high-power sub-sample ($\log(L_{1.4 GHz}) \ge 24.5$) and a low-power one ($24 \le \log(L_{1.4 GHz}) < 24.5$) were created, with the distinction between the two that roughly corresponds to the canonical division between FRI and FRII objects.

It was found that the significance in the environmental segregation signal is maintained only for low-power radio AGNs in the lowest and intermediate redshift bins, while for the high-power radio AGNs no significant signal is present. Therefore, higher overdensity richness enhance the probability that a galaxy hosts a low-power radio AGN. In conclusion, we found a clear correlation between radio AGN presence and environment up to $z \sim 2$, consistent with the scenario sketched in \cite{hickox09}.

\subsection*{Acknowledgements}
The authors acknowledge the financial contributions by grants ASI/INAF I/023/12/0 and PRIN MIUR 2010-2011 \textquotedblleft The dark Universe and the cosmic evolution of baryons: from current surveys to Euclid\textquotedblright.\\ 
\textcopyright$\:$ Springer International Publishing Switzerland 2016 \\
N.R. Napolitano et al. (eds.), \textit{The Universe of Digital Sky Surveys}, Astrophysics
and Space Science Proceedings 42, DOI 10.1007/978-3-319-19330-4\_17 \\

\end{document}